\documentclass[12pt,preprint]{aastex}
\usepackage{graphics,epsf}
\usepackage{amsmath}                
\usepackage{amsfonts}               
\usepackage{amssymb}                
\usepackage{epsfig}                 

\def \cm{~\rm{cm}}
\def \s{~\rm{s}}
\def \km{~\rm{km}}

\def \K{~\rm{K}}
\def \g{~\rm{g}}

\def \AU{~\rm{AU}}
\def \erg{~\rm{erg}}

\def \yr{~\rm{yr}}

\begin{document}

\title{POSSIBLE IMPLICATIONS OF THE PLANET ORBITING THE RED HORIZONTAL BRANCH STAR HIP 13044}

\author{Ealeal Bear\altaffilmark{1}, Noam Soker\altaffilmark{1}, and Amos Harpaz\altaffilmark{1}}

\altaffiltext{1}{Department of Physics, Technion -- Israel Institute of Technology, Haifa
32000, Israel; ealealbh@gmail.com; soker@physics.technion.ac.il}

\begin{abstract}
We propose a scenario to account for the surprising orbital properties of the planet orbiting the
metal poor red horizontal branch star HIP 13044.
The orbital period of 16.2 days implies that the planet went through a common envelope phase inside the
red giant branch (RGB) stellar progenitor of HIP 13044. The present properties of the star imply that the
star maintained a substantial envelope mass of $0.3 M_\odot$, raising the question of how
the planet survived the common envelope before the envelope itself was lost?
If such a planet enters the envelope of an RGB star, it is expected to spiral-in to the very inner region
within $\la 100 \yr$, and be evaporated or destructed by the core.
We speculate that the planet was engulfed by the star as a result of the core helium flash that caused this
metal poor star to swell by a factor of $\sim 3-4$. The evolution following the core helium flash is very rapid,
and some of the envelope is lost due to the interaction with the planet, and the rest of the envelope shrinks
within about a hundred years. This is about equal to the spiraling-in time, and
the planet survived.
\end{abstract}

\section{INTODUCTION}
\label{sec:intro}

In a recent paper \citet{Setiawan2010} announced the detection of
a planet orbiting the metal-poor red horizontal branch star (HB)
HIP 13044 (CD-36 1052) with an orbital period of $P=16.2 \pm 0.3~$days.
The star resembles red HB stars in globular clusters,
having an effective temperature, a mass, a radius, and a metallicity of
$T_{\rm eff} = 6025 \pm 63 \K$ (\citealt{Carney2008b};
\citealt{Roederer2010}), $M_\ast=0.8 \pm 0.1 M_\odot$ (\citealt{Setiawan2010}),
$R_\ast=6.7 \pm 0.3 R_\odot$ (\citealt{Carney2008b}), and ${\rm [Fe/H]} =-2.1$
(\citealt{Beers1990}; \citealt{Chiba2000}; \citealt{Carney2008b};
\citealt{Roederer2010}), respectively. From the perspective of
known exoplanets around main sequence stars, planets are rare
around metal poor stars (e.g., \citealt{Sozzetti2009}). The
detection of a planet around such a metal poor low mass star might
be taken as a surprise, although for more than a decade
theoretical studies have been proposing the presence of planets in
globular clusters (\citealt{Soker1998}; \citealt{SiessLivio1999};
\citealt{SokerHarpaz2000}, \citealt{SokerHarpaz2007};
\citealt{SokerHadar2001}; \citealt{SokerHersh2007}).

What we find surprising are the orbital semimajor axis of $a=0.116 \pm 0.01 \AU$ and its
eccentricity of $e=0.25 \pm 0.05$.
These parameters raise the following questions.
($i$) How come a companion of a mass of only $M_{\rm p} \simeq 7.5 M_{\rm Jup}$ survived a common
envelope phase with a RGB star that did not lose its entire envelope?
The properties of the star imply that the present envelope mass is $M_{\rm env} \sim 0.3 M_\odot$.
Naively, one would expect that an envelope that is $\sim 40$ times as massive as the planet
would have caused the planet to continue spiraling-in inside the progenitor RGB envelope within a very
short time ($\S \ref{sec:times}$).
($ii$) How come the orbit of a low mass companion that emerges from a common envelope is eccentric?
Gravitational drag and tidal interaction with the envelope are expected to circularize the orbit.

In section \ref{sec:angular} we show that the angular momentum of the system is about equal to
the initial (pre-RGB phase) orbital angular momentum of the planet. Therefore, there is no need to postulate
the presence of a third body in the system.
In section \ref{sec:times} we propose that the interaction of the planet with the RGB envelope
took place over a relatively short time of about a hundred years.
The interaction we speculate about in section \ref{sec:summary} was triggered by a brief but substantial
expansion of the star as a result of the core helium flash.
In section \ref{sec:summary} we also summarize and conclude that some globular clusters should be
a prime target for the search of planets around metal poor stars.

\section{ANGULAR MOMENTUM CONSIDERATIONS}
\label{sec:angular}

The possibility that planets spin-up RGB stars goes back to \citet{Peterson1983},
who try to account for the fast rotation of some HB stars.
Newer claims for planet-induced RGB stellar rotation include \citet{Soker1998}, \citet{Nelemans1998},
\citet{SiessLivio1999}; \citet{Reddy2002}, \citet{Denissenkov2004}, \citet{Massarotti2008},
\citet{Carney2003, Carney2008a} (who include HIP 13044),
and \citet{Carlberg2009, Carlberg2010}.
A systematic study was conducted by \citet{SokerHarpaz2000}, whose
calculations, assumptions, and approximations we adopt.

The parameters we use here are as given and derived by \citet{Setiawan2010}.
The companion mass is $M_{\rm p} = 1.25 \pm 0.05 M_{\rm Jup} / \sin i$, where $M_{\rm Jup}$ is Jupiter mass.
\citet{Setiawan2010} adopt the mean activity period to be due to rotation with
$P_{\rm rot} = 5.53 \pm 0.73~$days, and from that deduced the inclination angle of the orbital
plane to be $i=9.7^\circ \pm 1.3 $.
{{{  We note that, within the uncertainties, the ratio of orbital to rotation period equals 3.
The possibility of a tidal resonance should be examined. }}}
We will scale quantities with $i=10^\circ$, and hence with $M_{\rm p} \simeq 7.2 M_{\rm Jup}$.
As the stellar radius is $R_\ast=6.7 \pm 0.3 R_\odot$ (\citealt{Carney2008b}), the true rotation
velocity (on the equator) they derive is $v_{\rm rot} \sim 62 \km \s^{-1}$.
The orbital separation and eccentricity are $a_f=0.116 \pm 0.01 \AU$ and $e=0.25 \pm 0.05$, respectively.

The initial angular momentum is practically that of the planet at its pre-RGB orbit, as the stellar angular momentum is
negligible,
\begin{equation}
J_{\rm p0}=M_{\rm p} \left( GM_{\ast 0} a_0 \right)^{1/2}
= 8.5 \times 10^{50}
\left( \frac{M_{\rm p}}{7.2 M_{\rm Jup}} \right)
\left( \frac{M_{\ast 0}}{0.9 M_\odot} \right)^{1/2}
\left( \frac{a_0}{2\AU} \right)^{1/2} \g \cm^{2} \s^{-1},
\label{eq:Jp0}
\end{equation}
where $M_{\ast0}$ is the initial (pre-RGB) stellar mass, $a_0$ is the initial orbital separation,
and we assume a circular pre-RGB orbit.

The final angular momentum is carried by three components: The
planet in its final (eccentric) orbit, $J_{\rm p}$; the rotating
envelope, $J_{\rm env}$; and the mass that was expelled from the
star, $J_{\rm wind}$.  The present angular momentum of the planet is
\begin{equation}
J_{\rm p}=M_{\rm p} \left[ GM a_f (1-e^2) \right]^{1/2}
= 1.8 \times 10^{50} \left( \frac{M_{\rm p}}{7.2 M_{\rm Jup}} \right) \left(
\frac{M_\ast}{0.8 M_\odot} \right)^{1/2} \left( \frac{a}{0.116\AU}
\right)^{1/2} \g \cm^{2} \s^{-1},
\label{eq:Jp}
\end{equation}
where in the second equality we have substituted $e=0.25$ and
$a=0.116 \AU$ \citep{Setiawan2010}. The current stellar mass
is derived by taking a core mass of $0.5 M_\odot$ and an envelope
mass of $M_{\rm HBenv} \simeq 0.3 M_\odot$. The envelope and core
masses are estimated based on the results of \citet{Dorman1993}
and \citet{Dcruz1996}. Although the envelope mass is estimated to
be a little below $0.3 M_\odot$, the large uncertainties
concerning the angular momentum evolution justify using  $M_{\rm HBenv} = 0.3 M_\odot$.

The angular momentum of the rotating HB envelope is given by
\begin{equation}
J_{\rm env}=\alpha M_{\rm HBenv} R_\ast v_{\rm rot}=
1.7 \times 10^{49}
\left( \frac{M_{\rm HBenv}}{0.3 M_\odot} \right) \g \cm^2 \s^{-1},
\label{eq:Jenv}
\end{equation}
where we took $\alpha=0.01$ from \citet{Sills2000}, and we have substituted
$R_\ast =6.7 R_\odot$ and $v_{\rm rot}=62 \km \s^{-1}$ as given by \citet{Setiawan2010}.

To estimate the angular momentum carried by the wind we follow
\citet{SokerHarpaz2000} and assume that all the angular momentum
was deposited on the RGB, and all mass-loss took place after the
angular momentum was deposited. In the present scenario we
propose, most of the mass-loss process took place while the planet
was depositing its orbital angular momentum to the envelope. Such
a process reduces the angular momentum carried by the wind.
On the other hand, the same scenario implies that as the planet deposited
its angular momentum to the outer regions of the envelope, such that there
was not enough time for convection to redistribute the angular
momentum in the envelope. Such a process increased the angular
momentum carried by the wind. Over all there are large
uncertainties, and we use the approach of \citet{SokerHarpaz2000},
where more details are given.

We assume the ratio of the HB envelope mass to that of the initial
(before the interaction with the planet started) RGB envelope mass to be
$M_{\rm HBenv}/M_{\rm RGBenv} \simeq 0.65$.
For example, this is the ratio for an RGB stellar mass of $0.93 M_\odot$ and a core mass of $0.49M_\odot$
that give $M_{\rm RGBenv}=0.44M_\odot$, and for a present envelope mass of $M_{\rm HBenv} = 0.28M_\odot$.
The total angular momentum of the envelope after deposition by the planet and before
mass-loss have started, based on the present angular momentum of the envelope, is \citep{SokerHarpaz2000}
\begin{equation}
J_{\rm envR} \simeq J_{\rm env}\left(\frac{M_{\rm HBenv}}{M_{\rm RGBenv}}\right)^{-\delta}
=2.3 \times 10^{50}
\left( \frac{M_{\rm HBenv}}{0.3 M_\odot} \right)
\left( \frac{M_{\rm HBenv}/M_{\rm RGBenv}}{0.65} \right)^{6}
 \g \cm^2 \s^{-1},
\label{eq:Jwind}
\end{equation}
where the value of $J_{\rm env}$ was taken from equation (\ref{eq:Jenv}).
The angular momentum carried by the wind is $J_{\rm wind}=J_{\rm envR}-J_{\rm env}$.
The parameter $\delta$ is derived by \citet{SokerHarpaz2000}, which considered the range $ 4 \le \delta \le 7$;
we take here $\delta=6$, but the uncertainties should be kept in mind.

The total angular momentum carried by the different components after the planet started
depositing its angular momentum to the envelope is not much below the estimated initial
orbital angular momentum.
\begin{equation}
J_{\rm env}+J_{\rm wind}+J_{\rm p} \simeq 4 \times 10^{50} \g \cm^2 \s^{-1} \la J_{\rm p0}.
\label{eq:AngCons}
\end{equation}
Given the uncertainties, we can safely conclude that the total angular momentum of the system is about
equal to the initial orbital angular momentum of the planet.
This suggests that if a third body was present in the system its angular momentum was
small relative to the initial angular momentum of the observed planet.
Namely, it was a low mass planet and/or much closer to the star, and was swallowed earlier in the evolution.
In any case, it did not play a dynamical role in the interaction between the star and the observed planet
during the star transition from the RGB to the HB.
This important conclusion will be used in section \ref{sec:summary}.

\section{TIMESCALES CONSIDERATIONS}
\label{sec:times}

We now show that a secular (regular) RGB evolution with a planet around it
cannot lead to the present status of HIP 13044.
The planet starts its journey towards the RGB star when tidal interaction becomes strong enough
to reduce the spiraling-in time below the remaining stellar evolution time on the RGB.
For a planet mass of $M_p \simeq 0.01 M_{\ast 0}$ this occurs when the RGB radius reaches
a value of $R_{\rm RGB} \simeq 0.25-0.4 a_0$ (\citealt{Soker1996} eq. 6 there; \citealt{Villaver2009}; \citealt{Nordhaus2010}).
After the spiral-in process has started, the process is accelerated tremendously.
For a typical stellar parameters of a low mass star on the tip of the RGB,
the spiraling in time due to tidal interaction when the planet is outside the
envelope is (\citealt{Soker1996} and \citealt{Villaver2009}, where the weak dependance on the
other stellar parameters can be found)
\begin{equation}
\tau_{\rm in} \simeq 10^6 \left( \frac{M_{\rm p}}{0.01 M_{\ast 0}}
\right)^{-1} \left( \frac{a}{4 R_{\rm RGB}} \right)^8 \yr .
\label{eq:tidal1}
\end{equation}
When the planet reaches the RGB stellar surface the spiraling-in time is
$\tau_s \equiv \tau_{\rm in} (a={R_{\rm RGB}}) \simeq 10 \yr$.
After the planet enters deep into the envelope, gravitational drag will accelerate the spiraling-in
process, and the spiraling-in time becomes $\tau_{\rm in}(a<R_\ast) < 1 \yr$ \citep{Nordhaus2006}.

For a planet to end at an orbital separation of $a_f=0.116 \AU$ the envelope should
shrink to be less than $a_f$ within a timescale of $\tau_s \sim 10 \yr$.
This is $\sim 10$ times the dynamical time scale of an RGB star, and shorter than the thermal time
scale of the envelope $\tau_{\rm th-env} \equiv G M_{\rm RGB} M_{\rm env} /(R_{RGB} L_{\rm ASGB}) \simeq 100 \yr$, where
$R_{RGB}=0.5 \AU$ and $L_{\rm ASGB}=1000L_\odot$.

It is unlikely that the interaction of the planet with the envelope by itself can cause the envelope
to shrink over such a short time from $\sim 0.5 \AU$ to $< 0.1 \AU$.
If the core does not change the star stays a RGB star, that with an
envelope mass of $\sim 0.3M_\odot$ has a radius much larger than $0.1 \AU$.
The conclusion is that both the RGB core and envelope must be vigorously perturbed over a dynamical time scale
while the planet is spiraling in.
An interaction over a short time scale can account for the eccentric orbit of the planet in HIP 13044 as well,
as tidal and/or drag interaction over many orbits will circularize the orbit.

\section{DISCUSSION AND SUMMARY}
\label{sec:summary}

In section \ref{sec:angular} we concluded, based on angular momentum considerations,
that no tertiary object more massive than the planet was interacting with the RGB progenitor of HIP 13044,
unless it had a much smaller orbital radius.
In section \ref{sec:times} we argued that the interaction of the planet with the envelope must
have been on a very short time scale of $\la 100 \yr$.
We here present a speculative suggestion to account for these two conclusions.

We speculate that the core helium flash caused the envelope of the RGB progenitor to
expand by a factor of $\sim 3-4$ for a period of $\sim 100 \yr$.
In this scenario the planet was orbiting the RGB progenitor at an orbital separation of $a_0 \simeq 2-4 \AU$.
This speculative brief and large expansion of the RGB star during the core helium flash
{{{  is composed of two steps. In the first step a small fraction of the energy released by
 the hydrogen that is ignited in the outer parts of the core, is transferred outside the core.
This process is hard to study as it requires sophisticated 3D
numerical study of the core helium flash. This energy deposition
}}} was not found until now in numerical simulations of core helium flash
{{{  (e.g., \citealt{Mocak2011}). }}}
Despite that, we here present some arguments that should motivate future studies to look for such an effect,
{{{  in particular in rotating cores (that might have been spun up by an inner planet early on the RGB). }}}
 In that respect we note the comment made by \citet{Mocak2008} that many
processes following the core helium flash have some known
inconsistency that indicate that the core helium flash is not
fully understood. {{{  In the second step the energy deposited
at the base of the envelope causes the envelope its large and
brief expansion. Below we show that this is indeed the case. }}}

Along the RGB the star is powered by a hydrogen burning shell surrounding the almost pure helium core.
When a temperature of little over $10^8K$ is reached in the core, helium is ignited explosively.
Because of neutrino cooling prior to the ignition, the ignition itself occurs off-center (e.g., \citealt{Mocak2009}).
This core helium flash releases a vast amount of energy that cannot be carried radiatively, and thus convection
is triggered in the core.
It is thought that in most cases the He-burning and convective region reach up to the Hydrogen shell,
because the H-burning shell provides an entropy barrier against mixing
(\citealt{Campbell2010} and references therein).
\citet{Campbell2010} point out that in solar-mass RGB stars of primordial or hyper-metal-poor ([Fe/H]$ \leq -5.0$),
mixing might occur after all.
There are two reasons for this mixing \citep{Campbell2010}.
First, the core helium flash starts much farther away from the center in these low metallicity
models than in solar metallicity stars.
Second, the entropy barrier at the H-shell is much weaker in stars of very low metallicity because
the H-burning shell almost switches off at this stage of evolution (\citealt{Fujimoto1990}).
As a result of this mixing caused by the violent core helium flash in low metallicity stars
(\citealt{Campbell2008}; \citealt{Suda2010} for $[Fe/H] < -2.5$),
ignition of large amount of hydrogen occurs in these RGB stars
(\citealt{Mocak2008}; \citealt{Mocak2009}; \citealt{Mocak2010}).

\citet{Mocak2010} present a calculation of a core helium flash
followed by hydrogen ignition. Over the first year the hydrogen
burning provide $\sim 1 \times 10^{48} \erg$ (see their fig. 1).
After a year the hydrogen burning luminosity is $L_H \sim 10^6
L_\odot$. The huge energy production by the hydrogen burning
\citep{Mocak2010} and the core convection \citep{Blocker(1999)}
decay over a time scale of $\sim 10-100$ years. Most of this
energy stays in the core, and causes the core to swell.
{{{ We now show that it is sufficient that $\sim 5\%-10\%$ of the energy
released by the hydrogen burning leaks to the envelope to cause a substantial envelope expansion.

We run a spherical evolutionary stellar code based on the one used by \citet{Harpaz1981}
with updated opacities. The initial composition used is X=0.689, Y=0.31, and Z=0.001.
The code is spherical, and cannot follow the mixing of hydrogen to the core or the deposition of energy
from the core to the envelope, as these processes are highly non-spherical.
At the tip of the RGB, just as helium ignition starts, we manually add an anergy of $8.5 \times 10^{46} \erg$,
at the bottom of the envelope, just above the hydrogen burning shell.
This is $\sim 7\%$ of the energy released from the hydrogen burning reported  by \citet{Mocak2010}.
The duration of the energy injection was 7 years at a power of $L_{\rm in} =10^5 L_\odot$.
The initial model is presented in Figure \ref{fig:structure1}, while the model at the end of the
manually energy injection phase is shown in Figure \ref{fig:structure2}.
In Figure \ref{fig:radii} we show the evolution of the outer radius and the outer boundary of the convective region.
For the tidal interaction the outer boundary of the convective region is important.
In this calculation the outer radius of the convective region increases by a factor of $\sim 4$.
After $\sim 100 \yr$ the star shrinks back to its original radius.
This calculation does not include mass loss, that can lead to further envelope contraction.
\begin{figure}  
\includegraphics[scale=0.6]{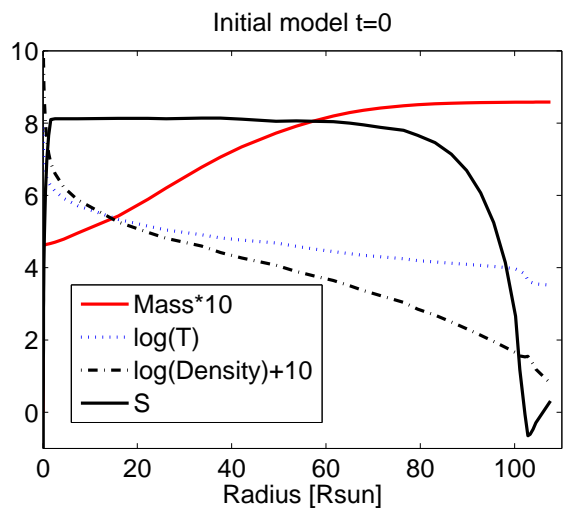}
\caption{\footnotesize{
The structure of the envelope just before the addition of energy at the base of the envelope
(above the hydrogen burning shell). Shown are the mass ($M_\odot$), temperature ($K$) and density ($\g \cm^{-3}$),
and the entropy (relative units). Convective regions are where the entropy decreases outward.
 }}
\label{fig:structure1}
\end{figure}
\begin{figure}
      \includegraphics[scale=0.6]{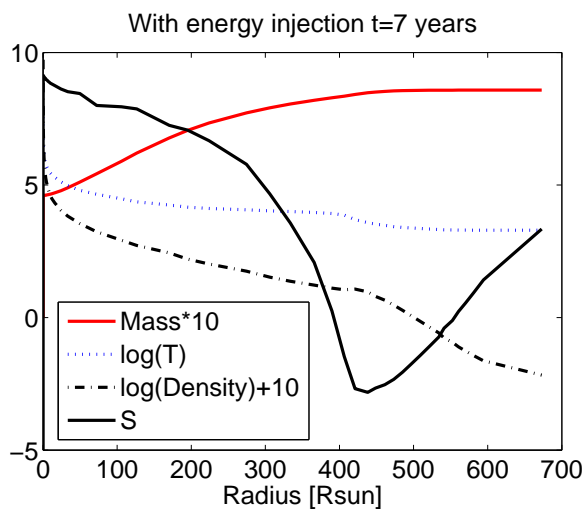}
\caption{\footnotesize{
Like Figure 1, but after 7 years of energy injection at the base of
the envelope at a power of $L_{\rm in} =10^5 L_\odot$.
 }}
\label{fig:structure2}
\end{figure}
\begin{figure}
   \includegraphics[scale=0.4]{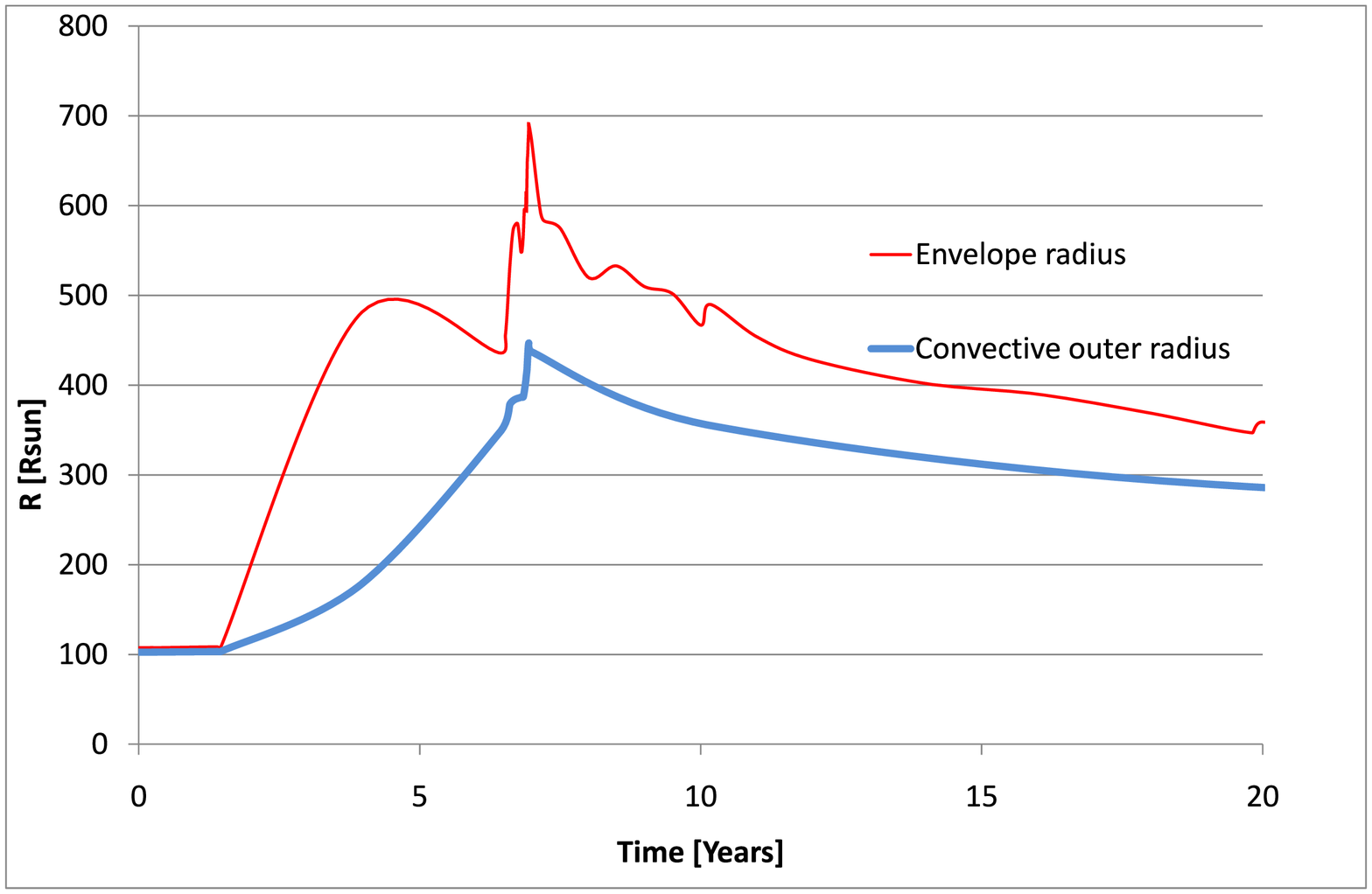}
   \vskip -2.0cm
   \includegraphics[scale=0.4]{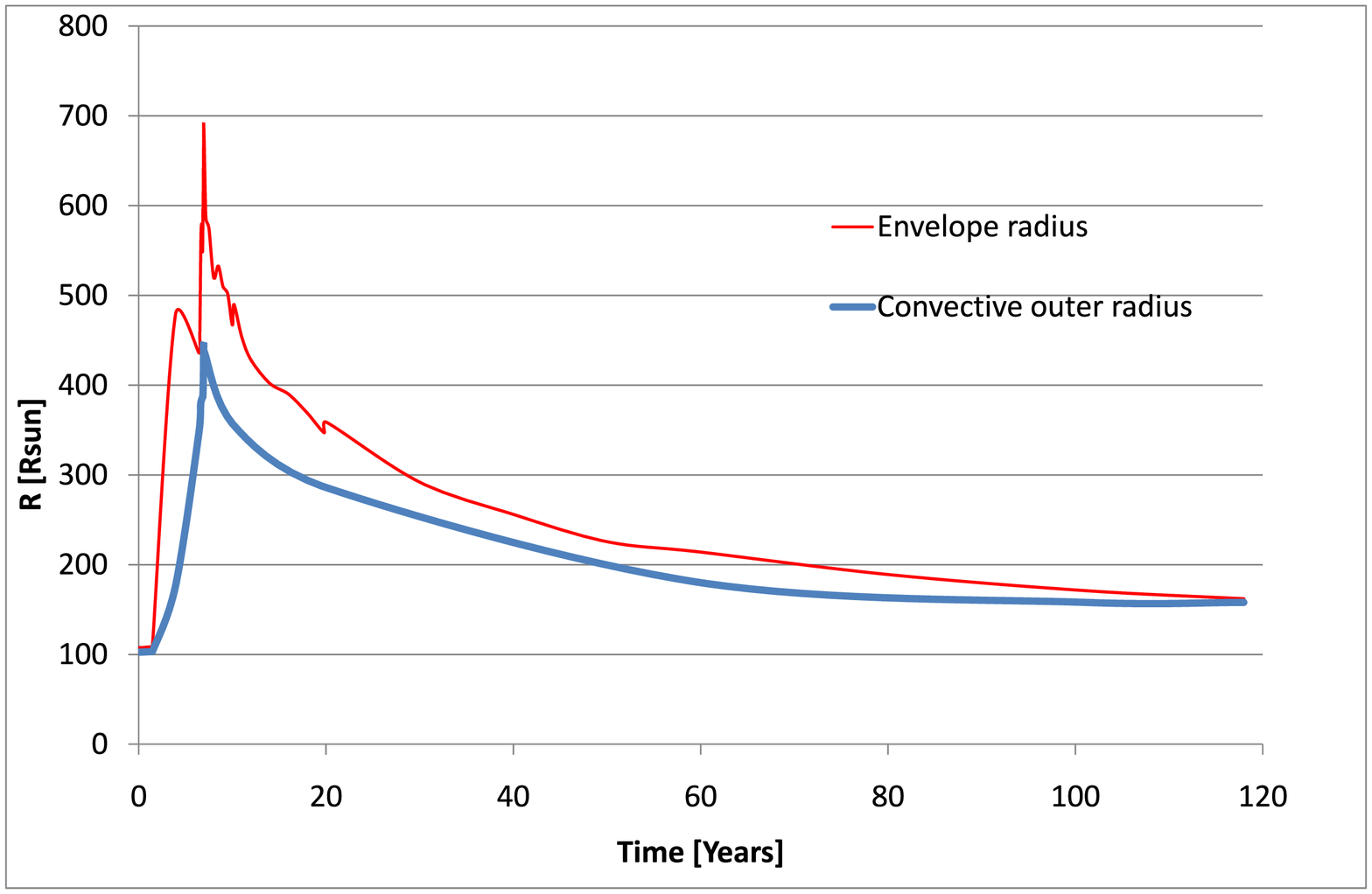}
\caption{\footnotesize{
The outer radius of the envelope (thin red line) and the outer boundary of the convective region
(thick blue line) as function of time. Upper and lower panels show the same calculation but for
different time spans.
The noise in the graph (wiggling of the lines) demonstrates the numerical limitation of the code.
It has a finite number of numerical shells, and the rapid increase in radius is done out of thermal equilibrium.
These effects cause the noise.  }}
\label{fig:radii}
\end{figure}
}}}

The spiraling in of the planet with the properties of the HIP 13044 system when the inflated envelope reaches
the planet orbit is $\sim 10-100 \yr$ (\citealt{Soker1996}). The planet starts spiraling-in on a time scale about equal
to that of the inflated envelope.

We suggest that during the core helium flash of low metallicity RGB stars, a small fraction
(few percents) of the energy liberated by the hydrogen burning, that takes place in the outer regions of the core,
is transferred to the envelope. As we showed above, this causes the outer region of the envelope to substantially expand.
A substantial increase in radius is found in some calculations of shell helium flashes (thermal pulses)
in AGB stars (e.g., \citealt{Schlattl2001}; \citealt{Boothroyd1988}).

As the planet spirals-in in such an inflated envelope, it enhances the mass loss rate by
depositing gravitational energy and by spinning-up the envelope \citep{Soker2004}.
The calculation of the mass loss is complicated and beyond the scope of this paper, but
is expected to be of the order of $\sim 0.1 M_\odot$ based on the properties of HIP 13044.
The rest of the envelope, in our scenario, shrinks below the orbital separation of the spiraling-in
planet before the planet manages to spiral-in below $\sim 0.1 \AU$,
and the spiraling-in ceases before the envelope is lost.
This envelope contraction is caused by the changes in the core following the core helium flash,
and it is expedited by the rapid mass loss caused by the spiraling-in planet.
The rapidly changing envelope properties imply a rapidly varying tidal interaction, that instead of
circularizing the envelope causes the eccentricity to increase.
The overall evolution lasts for several dynamical time scales.
This is possible because of the energy that is transferred from the core flash to the envelope over
a very short time scale. This is crucial for our proposed scenario to work.

We expect only low metallicity low-mass stars (Pop II) to experience the $\sim 100 \yr$ long
inflated envelope phase. The total number of such objects in all globular clusters is expected to be $<1$.
Therefore, it will be extremely hard to find such stars in globular clusters and in the field.
Even if found, they can be easily confused with AGB stars, unless they are followed for tens of years.
In any case, the planet orbiting the red HB star HIP 13044 shows that planets can exist in globular clusters,
and they can influence the evolution of the star. In particular they can increase the mass-loss rate and
lead to the formation of blue HB stars \citep{Soker1998}.
We therefore suggest that globular clusters with a large population
of blue HB stars be a prime target for exoplanet research.

Another possible process that might have occurred in this system is that it started as a multi-planet system
\citep{Bear2010}. There was an inner planet that was engulfed earlier on the RGB.
This planet spiralled inward all the way to the core vicinity, and might have spun-up the core.
Future 3D simulations of core helium flashes should check whether core rotation can facilitate the
transfer of energy from the flashing core to the envelope.

The Research was supported in part by the N. Haar and R. Zinn Research fund at the Technion,
the Israel Science Foundation, and the Center for Absorption in Science, Ministry of Immigrant Absorption,
State of Israel.


\label{lastpage}

\end{document}